# Generation and Focusing of Orbital Angular Momentum Based on Polarized Reflectarray at Microwave Frequency

Fengxia Li, Haiyan Chen, Yang Zhou, Jian Wei You, *Member, IEEE*, Nicolae C. Panoiu, *Member, IEEE*, Peiheng Zhou, *Member, IEEE*, and Longjiang Deng

*Abstract*—A novel polarized reflectarray is designed, fabricated, and experimentally characterized to show its flexibility and efficiency to control wave generation and focusing of OAM vortices with desirable OAM modes in the microwave frequency regime. In order to rigorously study the generation and focusing of OAM, a versatile analytical theory is proposed to theoretically study the compensation phase of reflectarray. Two prototypes of microwave reflectarrays are fabricated and experimentally characterized at 12 GHz, one for generation and one for focusing of OAM-carrying beams. Compared with the OAM-generating reflectarray, the reflectarray for focusing OAM vortex can significantly reduce the beam diameter, and this can further improve the transmission efficiency of the OAM vortex beams. We also show that the numerical and experimental results agree very well. The proposed design method and reflectarrays may spur the development of new efficient approaches to generate and focus OAM vortex waves for applications to microwave wireless communications.

## I. Introduction

With the rapid development of wireless communication technology, it has become imperative to improve the system capacity and spectral efficiency to meet the exponential growth of demand in the area of broadband data transfer, data centers, and cloud-based services. To overcome this challenge, orbital angular momentum (OAM) has emerged as an effective means to carry information due to its multiple orthogonal modes can simultaneously transfer at the same frequency in a single communication channel, offering alternative and flexible degrees of freedom [1] without increasing the frequency bandwidth [2]-[4]. As a result, the use of OAM in wireless communications has become a field of intense research, owing to its potential applications, especially in the microwave regime [5]-[8]. Therefore, it is very important to develop effective methods to generate OAM vortex waves at microwave frequency. Within the broad array of applications of OAM, how to efficiently generate, manipulate and receive OAM-carrying beams are the most important issues. Recently, versatile antennas and antenna arrays have been demonstrated to generate OAM beams, the underlying principle being the introduction of an azimuthal phase factor in order to generate single or multiple OAM modes [9]-[11]. However, such devices have some inherent limitations. Thus, in practical applications, antennas based on spiral reflectors have large thicknesses and are usually difficult to fabricate [12], [13]. Furthermore, to generate OAM using antenna arrays one needs a complex feeding network, so as to generate a phase difference between array elements [14]-[16]. This can increase the complexity of system integration and production costs. However, instead of the above designs, reflectarray are gradually considered to be one of the practical ways for generating vortex waves in the radio frequency domain. Reflectarray can be designed to produce single or multiple beams using single or multiple layer configuration at different frequency bands [17]-[20]. A single-layer, dual-frequency unit for generating OAM in the microwave range for multifunctional OAM with required OAM mode, beam number, and direction was reported in [21]. According to [22], a high-efficiency planar reflectarray with a small size was proposed to effectively convert arbitrary polarized waves to OAM waves. Further, Jiang *et al.* [23] designed a class of low-profile and broadband dual-circularly polarized reflectarrays with independent beamforming for circularly polarized waves with opposite handedness. Meanwhile, a dual-band dual-polarized reflectarray for generating dual beams with respect to carrying two different OAM topological charges operating in the C-band in horizontal polarization and in the X-band in vertical polarization was proposed in [24]. They main concern with high efficiency and high OAM mode purity in a low-profile configuration. More recently, metasurfaces based on catenary and other structures also have been reported to generate vortex waves successfully, providing another novel route to achieve OAM beam.

Manuscript submitted for review July 28, 2020; revised September 3, 2020; accepted October 31, 2020. This work was supported in part by the National Natural Science Foundation of China (No. 51972046 and 51772042), in part by "111" Center (No. B13042), in part by China Scholarship Council (CSC), and in part by European Research Council (ERC-2014-CoG-648328). The work of experimental testing was supported in part by Professor Shiwei Qu, Dr. Yafei Wu and Dr. Yunpeng Zhang. (Corresponding author: Haiyan Chen. phone: +86(28)61830846; fax: +86(28)61830846; e-mail: chenhy@uestc.edu.cn )

F. X. Li, H. Y. Chen, Y. Zhou, P. H. Zhou, L. J. Deng are with National Engineering Research Center of Electromagnetic Radiation Control Materials, Key Laboratory of Multi-spectral Absorbing Materials and Structures of Ministry of Education, State Key Laboratory of Electronic Thin Film and Integrated Device, University of Electronic Science and Technology of China, Chengdu, 610054, China. (e-mail: lfx0519lfx@163.com, chenhy@uestc.edu.cn, zhouyaug@163.com, zhouph1981@163.com, denglj@uestc.edu.cn).

F. X. Li, J. W. You and N. C. Panoiu are with Department of Electronic and Electrical Engineering, University College London, Torrington Place, London WC1E 7JE, United Kingdom (e-mail: lfx0519lfx@163.com, j.you@ucl.ac.uk, n.panoiu@ucl.ac.uk).

Metasurfaces have been demonstrated to have enhanced ability to facilitate OAM generation and processing [25]-[27]. As the two-dimensional equivalent of metamaterials, metasurfaces have been widely used to manipulate electromagnetic (EM) waves [28]-[32], chiefly due to their ultra-thin and sub-wavelength characteristics. In particular, due to their remarkable abilities to manipulate the amplitude, phase and polarization state of EM waves [33]-[37], it is possible to employ metasurfaces to generate vortex beams with an arbitrary topological number, as it has been confirmed by a series of experiments in the optical and microwave regime [38]-[43]. Yu *et al.* [44] demonstrated a reflective metasurface to generate OAM in radio frequency domain at a single frequency point, but the bandwidth is narrow and the size is relatively large. A metasurfaces composed of Pancharatnam-Berry (PB) phase elements are demonstrated to control the conical beam generation by Ding *et al.* [45], however, it did not consider the issue of transmission and efficiency. Apart from this, OAM vortex waves have annular structure and are divergent, thus the inner diameter of the beam is a key factor that determines their suitability for communications based on specific OAM modes. In addition, as the order of the OAM-mode and the propagation distance increase, the degree of beam divergence increases, too. Therefore, it is particularly important to develop effective methods for focusing of OAM vortex waves.

In the microwave regime, despite previous research focused on OAM beam generation, it has not been demonstrated that a reflectarray based on reflective polarized converter can be used to generate and focus OAM beams simultaneously. Recently, the control of polarization states of EM waves based on periodic structure has attracted intense research interest at the frontier of science and engineering due to the ability to modify the amplitude and phase of EM waves [46]-[48]. By controlling the phase and amplitude responses of the unit-cells in reflectarray, a vortex beam with arbitrary topological charge can be easily realized. This has spurred great interest in designing polarized reflectarrays to generate and focus OAM beams. In this paper, we propose two types of polarized reflectarrays to generate and focus OAM beams at microwave frequencies. The cross-polarization conversion efficiency of these polarized elements is over 70% in the bandwidth above 55%, and these unit cell can achieve $2\pi$ phase variation. These polarized reflectarrays provide great flexibility in phase control while greatly suppressing the transmission loss. Moreover, a theoretical formula for the phase distribution is derived and used to design the two reflectarrays for the generation and focusing OAM beams. These reflective polarized reflectarrays have the advantages of being ultra-thin and light-weight, and provide great flexibility in tailoring the phase distribution in its unit cell. Two of these reflectarrays were designed, fabricated, and measured at 12 GHz, so as to validate the theoretical predictions, generation and focusing of OAM vortex waves is achieved by using quasi-spherical beams and we found that the results of numerical simulations agree very well with the experimental measurements. Thus, the simulations and measurements demonstrate that the generation and focusing of OAM-carrying beams can be flexibly achieved by using polarized reflectarrays. However, considering the communication requirements, it can be found that it is necessary to further increase the transmission and focusing distance of the OAM beam, which will also be one of our future research topics.

## II. REFLECTARRAYS DESIGN AND RESULTS

A reflective polarized converter based on a single-layer unit cell is chosen here. The proposed element is sub-wavelength, and its parameters are illustrated in Fig. 1(a). The square split ring and the metal ground plane are separated by a Teflon dielectric spacer, and the reflection phase varies with the change of parameters *b* and *w*. The unit cell period $p = 10$ mm, thickness of substrate $h = 3$ mm with relative permittivity $\varepsilon_r = 2.65$, and the other geometrical parameters of the element are: $a = 6$ mm, $a_1 = 5$ mm and $b_1 = 3$ mm. The values of parameters and reflection phase of the element are listed in Table I.

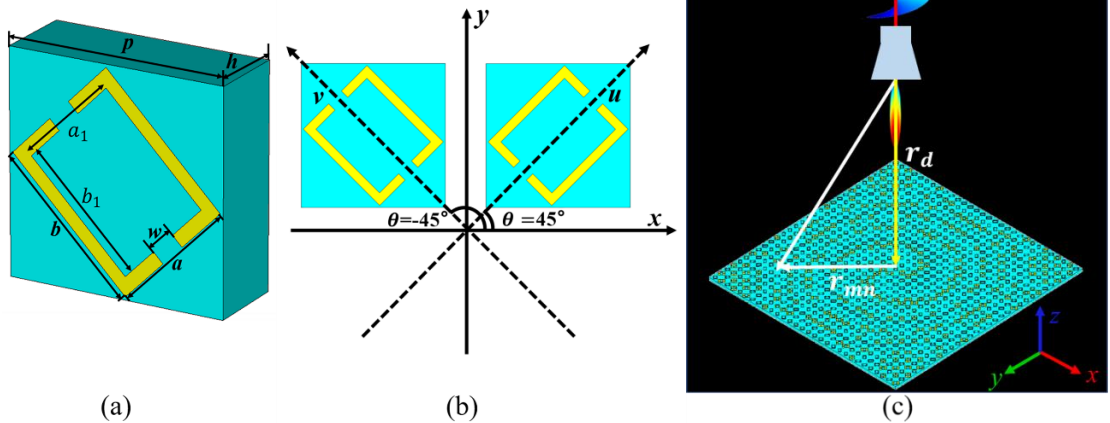

Fig. 1 (a) Schematic illustration of the reflective polarized converter with variable *w* and *b*. (b) Polarization conversion structures of 45° symmetrical and -45° symmetrical, and $\theta$ is the angle between the *x*-axis and the *u*/*v* symmetry axis of the structure. (c) Configuration of OAM-generation reflectarray.

TABLE I
PARAMETERS OF SUB-WAVELENGTH ELEMENTS

| Parameters | Element1 | Element2 | Element3 | Element4 | Element5 | Element6 | Element7 | Element8 |
|---|---|---|---|---|---|---|---|---|
| $b$/mm | 3.6 | 7.8 | 5.0 | 4.4 | 3.6 | 7.8 | 5.0 | 4.4 |
| $w$/mm | 0.8 | 0.2 | 0.2 | 1.0 | 0.8 | 0.2 | 0.2 | 1.0 |
| $\theta$/° | -45 | 45 | 45 | 45 | 45 | -45 | -45 | -45 |
| Unit cells | | | | | | | | |
| Reflection Phase /° | -172 | -116 | -74 | -32 | 9 | -296 | -255 | -212 |

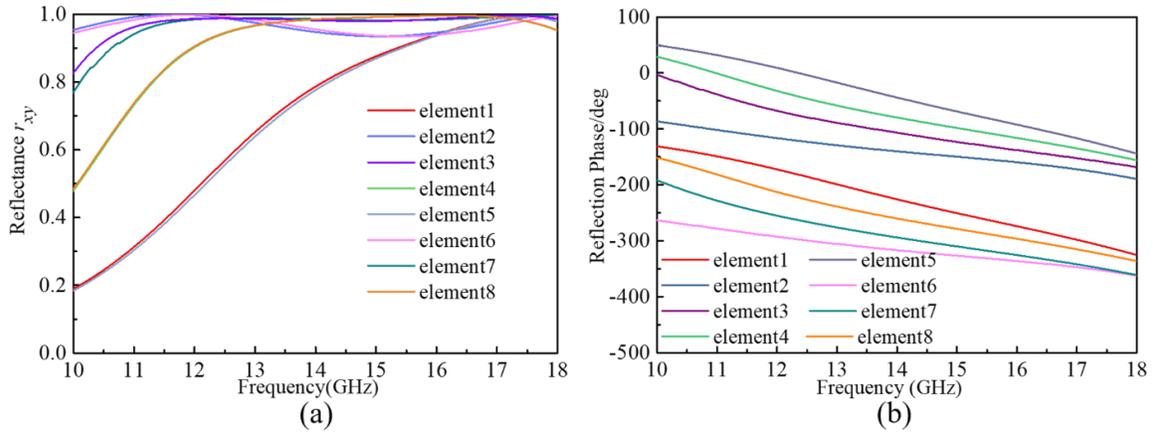

Fig. 2 (a) Simulation of cross-polarization reflection coefficient of the polarized unit cells at 12 GHz versus frequency with the change of parameters $b$ and $w$. (b) Simulated reflection phase of the polarized unit cells at 12 GHz versus frequency with the change of parameters $b$ and $w$.

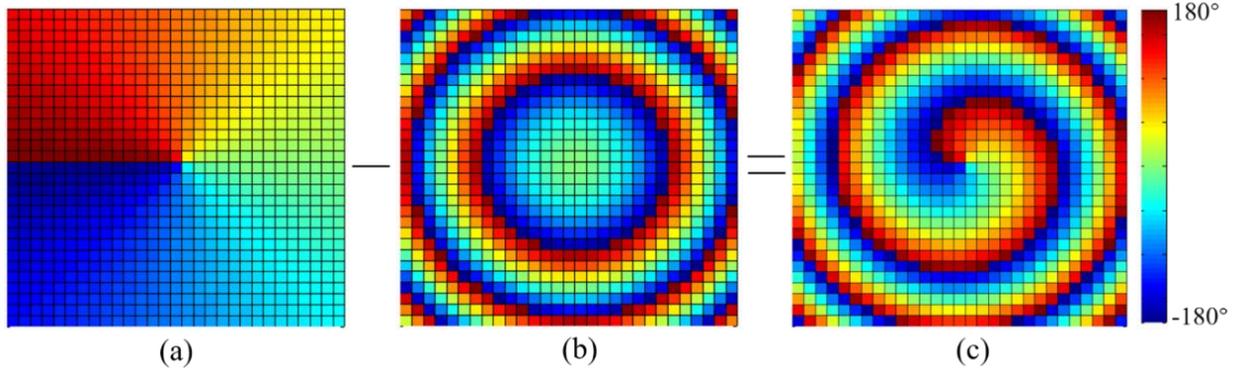

Fig. 3 Calculation process and phase distribution compensated on the reflectarray based on the polarized reflective elements. (a) Output phase of vortex wave. (b) Input phase of incident source. (c) Compensate phase of reflectarray.

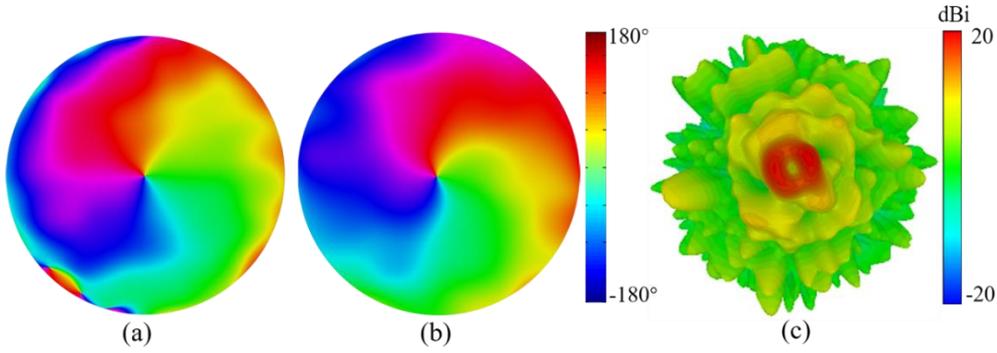

Fig. 4 Simulation results of generating OAM vortex waves with topological charge $l = 1$ at 12 GHz. (a) Phase distribution of electrical field at cross sections $z = 600$ mm ($z = 24\ \lambda_0$) in x-y plane. (b) Phase distribution of electrical field at cross sections $z = 800$ mm ($z = 32\ \lambda_0$) in x-y plane. (c) Top view of simulated radiation pattern in the far-field zone.

Under linearly polarized incidence, the cross-polarization reflective coefficient is above 70% with the bandwidth of more than 55%, as show in Fig. 2(a). In addition, the results plotted in Fig. 2(b) demonstrate that the phase of the reflective cross-polarized components can change by 360° by varying the values of b and w, which is enough to produce OAM vortex waves with arbitrary topological number.

The proposed reflectarray consists of polarized elements, metallic ground plate, and an illuminating feed antenna, as illustrated in Fig. 1(c). Considering a reflectarray, which consists of M*N elements that are illuminated by a feed source located at the position of $r_d$, $r_{mn}$ is the distance between the mnth element and the coordinate center, and the generating OAM vortex waves propagate along the z-axis.

## III. ANALYSIS AND DISCUSSION

### A. Numerical Results

According to the Helmholtz equation in free space [49], for vortex waves that propagate along the z-axis, the vector electric field $\boldsymbol{E_l}$ in cylindrical coordinates can be expressed as follows

$$\boldsymbol{E_l}(r,\varphi,z) = \boldsymbol{E_0}\exp(il\varphi)\exp(-ik_0 z) \quad (1)$$

where $l$ is an arbitrary integer representing the topological charge, and is also the OAM mode number. We assumed that if the coordinate of a unit cell is $(x, y)$, then $\varphi = \arctan(y/x)$, which is related to the origin point, $k_0$ is the wave number of free spaces, and $\boldsymbol{E_0}$ is a constant vector. As illustrated in Fig. 1(c), we can get the expression of an input phase of the standard horn, that is

$$\varphi_I = k_0\sqrt{x^2 + y^2 + r_d^2} \quad (2)$$

Then the compensating phase required at each polarized reflective unit cell in the desired direction can be obtained by

$$\varphi_{mn} = l\arctan(y/x) - k_0\sqrt{x^2 + y^2 + r_d^2} \quad (3)$$

where $\varphi_{mn}$ is the azimuthal angle of the mnth element and $r_d$ is the distance between the horn and the reflectarray. Based on the phase distribution of the required function of the reflectarray, the specific unit cell at any location can be designed.

Numerical results shown in Fig. 3 reveal that the phase variation of the reflectarray must cover 360°. The output phase of the vortex wave and the input phase of the incident source are shown in Fig. 3(a) and Fig. 3(b), respectively. The necessary compensating phase generated by the reflectarray can be calculated by using EM field superposition principle as given in (3), and it is given in Fig. 3(c). Theoretically, OAM vortex waves with arbitrary topological charges can be generated in terms of the continuous spatial phase change by employing reflectarrays with periodic elements.

### B. Simulation Results

For illustration, we designed a reflectarray using polygonal elements to generate OAM beams. The reflectarray contained 30×30 elements based on these eight basic unit cells presented in Table I. The layout is a square array with the dimension of 300 mm×300 mm, as shown in Fig. 1(c). The topological charge of OAM-generation design was chosen to be 1. In order to verify the performance of the designed reflectarray, numerical simulations are performed using the CST Microwave Studio for all these schemes with normal incident wave and polarization along the y-axis, and full-wave simulations based on the finite difference time-domain technique (FDTD) have been performed at 12 GHz ($\lambda_0 = 25$ mm), to analyze the characteristics of OAM-generation by the reflectarray. The unit cell boundary conditions were applied along the x/y-directions. A horn antenna was used as the feeding source at normal incidence, and the feed point was located in front of the reflectarray at $10\ \lambda_0$ distance away from the reflectarray along the z-axis. Figs. 4(a)-(b) show the simulated phase distribution at the cross sections $z = 600$ mm ($z = 24\ \lambda_0$) and 800 mm ($z = 32\ \lambda_0$), which are perpendicular to the propagation direction, the characteristics of the vortex waves can be seen from the phase distribution. Fig. 6(a) and (b) show the simulated distribution of energy at the cross sections $z = 600$ mm and 800 mm, the inner diameters(D) of the energy distribution in x-y plane are $1.8\ \lambda_0$ for $z = 24\ \lambda_0$ and $3.0\ \lambda_0$ for $z = 32\lambda_0$, respectively. From these results, it can be clearly seen that the inner diameters of the energy distribution along the propagation direction increases when the distance increases. The corresponding simulation of the 3D scattering patterns in the far-field zone is shown in Fig. 4(c), and the results reveal that there is an amplitude null in the center of the beam, which agrees with the properties of OAM vortex beams with topological charge equal to 1. Fig. 6(c) illustrates the simulated distribution of energy in the *y-z* plane. It can be seen in this figure that, as expected, when the propagation distance increases, the radius of the beam increases, indicating that the vortex beam with OAM mode diverges due to diffraction.

As shown in Fig. 6(a) and (b), OAM vortex waves would expand along propagation direction, which creates serious problems at the receiver end in communication systems. Therefore, it is necessary to reduce the radius of the vortex beam along the propagation path. In order to achieve this, the phase function of the reflectarray should be equivalent to the interference pattern between a focusing phase factor and a spiral phase plate, so that the cross-polarized reflective converter can constructively focus OAM beams with arbitrary topological charge. This can be fulfilled by the superposition of the spiral phase profile and focusing phase factor [50]. The required compensation phase $\varphi_{mn}^c$ of a unit cell located at the position of $(x, y)$ can be expressed as:

$$\varphi_{mn}^c = l\arctan(y/x) + k_0(\sqrt{x^2+y^2+f_1^2}-|f_1|) - k_0\sqrt{x^2+y^2+r_d^2} \quad (4)$$

where $f_1$ represents the focal length.

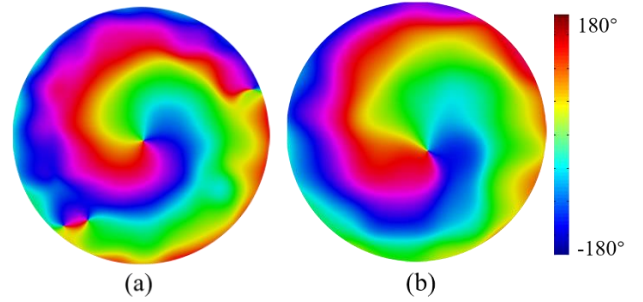

Fig. 5 Simulation phase distributions of electrical field for OAM-focusing vortex beam with topological charge $l = 1$ at 12 GHz in *x-y* plane: (a) $z = 600$ mm and (b) $z = 800$ mm.

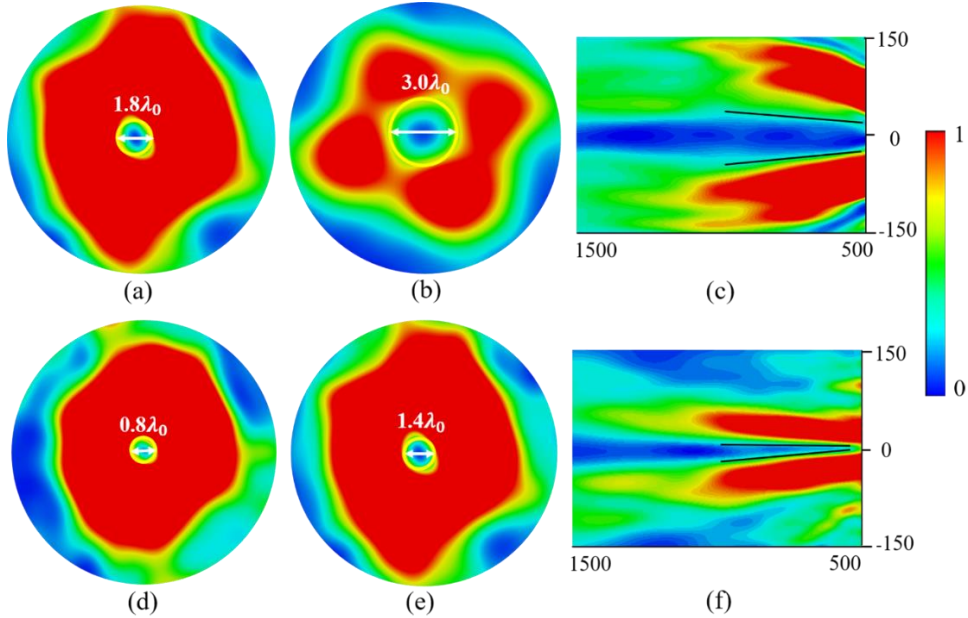

Fig. 6 Simulation results of electrical field energy of polarized reflectarrays for generating and focusing vortex beams with topological charge of 1 at 12 GHz. (a) Simulated distribution of energy in the *x-y* plane for OAM-generating reflectarray at cross sections $z = 600$ mm. (b) Simulated distribution of energy in the *x-y* plane for OAM-generating reflectarray at cross sections $z = 800$ mm. (c) Simulated distribution of energy for OAM-generating beams in the *y-z* plane. (d) Simulated distribution of energy in the *x-y* plane for OAM-focusing reflectarray at cross sections $z = 600$ mm. (e) Simulated distribution of energy in the *x-y* plane for OAM-focusing reflectarray at cross sections $z = 800$ mm. (f) Simulated distribution of energy in *y-z* plane for OAM-focusing beams.

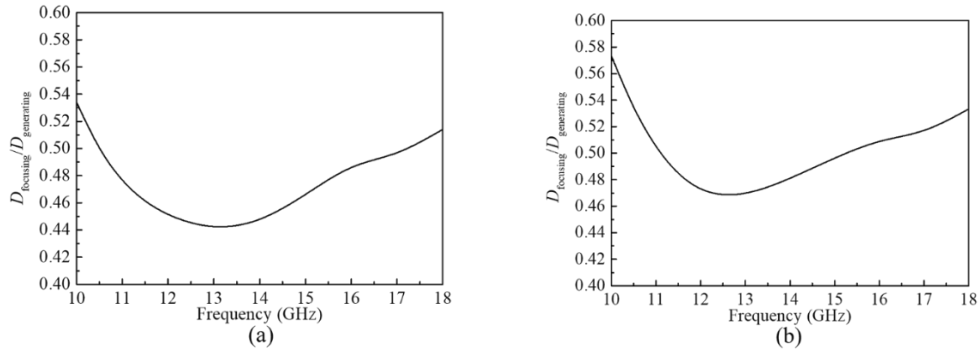

Fig. 7 Simulation beam diameters ratio of polarized reflectarray between OAM-focusing and generation vortex beams with topological charge of 1 versus frequency. (a) $z = 600$ mm and (b) $z = 800$ mm.

Simulations were performed at 12 GHz ($\lambda_0$ = 25 mm) and the topological charge of OAM-focusing design was chosen to be 1. The parameters of these eight unit cells are the same as those of the OAM-generation design. This focusing reflectarray, which is composed of 30×30 unit cells, has focal length of $24\lambda_0$ ($f_1$ = 600 mm). Fig. 5 shows the simulation results of phase distribution at different propagation distances in the x-y plane at 600 mm and 800 mm. It can be seen from these figures that, when compared to the phase at the same position corresponding to the OAM-generating reflectarray, the diameters of the vortex wave are reduced and the divergence of OAM beams is suppressed. In Fig. 6, which shows the spatial profile of the energy of the beam, it can be clearly observed that doughnut-shaped energy intensity patterns were obtained, and that the energy distribution approaches the center axis during propagation when compared with the OAM-generating reflectarray at cross sections z = 600 mm and 800 mm, which can be seen in Fig. 6(d) and (e). The inner diameters of energy map of OAM-generation reflectarray are $1.8\lambda_0$ and $3.0\ \lambda_0$ for z = $24\ \lambda_0$ and $32\ \lambda_0$ in x-y plane while for OAM-focusing reflectarray are $0.8\lambda_0$ and $1.4\ \lambda_0$, respectively. At the same propagation position, the diameters of OAM-focusing beam are reduced, the OAM-focusing reflectarray is obviously more constrictive, while the diameter of energy ring is about half of that OAM- generation reflectarray in the position of $24\lambda_0$ and $32\lambda_0$. which demonstrates the efficiency of the OAM-focusing of this reflectarray.

In Fig. 6(c) and (f), we present the simulated results of OAM- generating and focusing vortex beams in the y-z plane. From these figures, one can see that as the propagation distance increases, the radius of the OAM-focusing beams is reduced when compared to that of the OAM-generating beam, which indicates that the OAM-focusing reflectarray can efficiently arrest deleterious diffractive effects.

As is indicated by the above discussions, further numerical simulations reveal that the OAM-focusing reflectarray is obviously more constrictive, which can be seen in Fig. 7. As show in Fig. 7(a) and (b), the diameter of the OAM-focusing reflectarray is smaller than that of the OAM-generation reflective reflectarray at the distance of 600 mm and 800 mm, respectively. And the frequency points with uniform phase difference and larger cross-polarization conversion amplitude have a better focusing effect.

IV. FABRICATION AND EXPERIMENTAL RESULTS

To experimentally validate the design principle, theoretical method and simulation results, prototypes of the two OAM-generation and focusing reflectarrays are shown in Fig. 8(b) and (c) contains 30×30 unit cells and has an overall size of 300 mm × 300 mm, which were fabricated using low-cost printed circuit board (PCB) technique and measured in an anechoic chamber, as shown in Fig. 8. The near-field planar scanning technique was applied to characterize the OAM vortex waves. The experimental system is shown in Fig. 8(a).

We measured the scattering patterns of the samples in a standard microwave anechoic chamber and the measured results are calibrated to a same size metal plate. These measurements were performed at the frequency of 12 GHz ($\lambda_0$ = 25 mm) in x-y plane in order to get the electric field intensities and phase distributions. In the measurement system, a wideband horn antenna (80180HA20N) is employed as the excitation at a distance of d= 250 mm ($10\ \lambda_0$ ) away from the center of the reflectarrays. An active antenna (120WOEWPN) is used as field probe and is capable to measure the cross-polarization component of the reflected electric field through the network analyzer (8720ET). The horn antenna and probe were connected with the network analyzer by the coaxial cable to measure the sample. The probe shown in Fig. 8(a) is used as the receiving end whose position is controlled by a motion controller. The orientation of the probe antenna here is set to be vertical to the samples in order to obtain the components of the electric fields. In order to observe the stability of the vortex wave as it propagates, the scanning plane is set to be 600 mm and 800 mm from the samples. A small scanning step of 1 mm is selected to acquire intensities and phase distributions at the sampling planes. With the variation of the position of the field probe via the motion controller, the x-y plane can be covered and the experimental field intensities and phase profiles can be measured.

The measured energy intensity and phase distribution of the reflected electric field are shown in Fig. 9 and Fig. 10 in the x-y plane at cross sections z = 600 mm and z = 800 mm, respectively. Compared to the simulated OAM spatial field distribution (Fig. 4, Fig. 5 and Fig. 6), we can see that the OAM vortex wave of generation and focusing reflectarrays with l = 1 is observed. The simulated and measured results are in good agreement. There is a subtle difference in the measured energy intensity and phase distribution due to the supporting structure of the feeding horn and the coaxial cable of the reflection system. However, the major features of the spatial phase distribution can be clearly recognized. The typical feature of the field intensity is a perfect doughnut-shaped intensity map, and we can see clearly that there is a null in the center of the beam, which corresponds to the property of a vortex waves with OAM. From Figs. 9(a), (b) and Figs. 10(a), (b), it can be clearly seen that comparing to OAM mode at the same propagation distance, OAM-focused beam is more radially confined, and the radius of energy ring is smaller than the OAM-generation beam, and the measured inner diameters of the energy distribution for OAM-generation vortex waves are $1.8\ \lambda_0$ and $2.8\ \lambda_0$ for z = $24\ \lambda_0$ and $32\ \lambda_0$ in x-y plane, at the same position, the inner diameters of the energy distribution for OAM-focusing vortex waves are $0.9\ \lambda_0$ and $1.6\ \lambda_0$, respectively, which agrees with the simulation results. The phase distributions shown in Figs. 9(c), (d) and Figs. 10(c), (d) are also in agreement with the simulation results, which validates our theoretical approach.

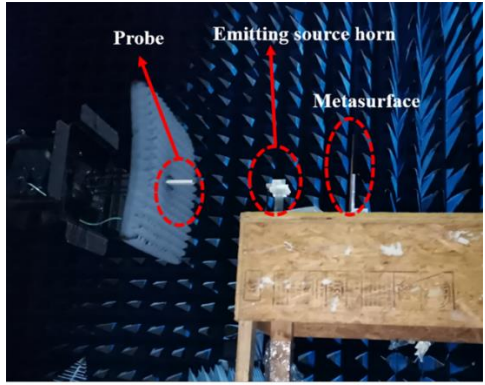

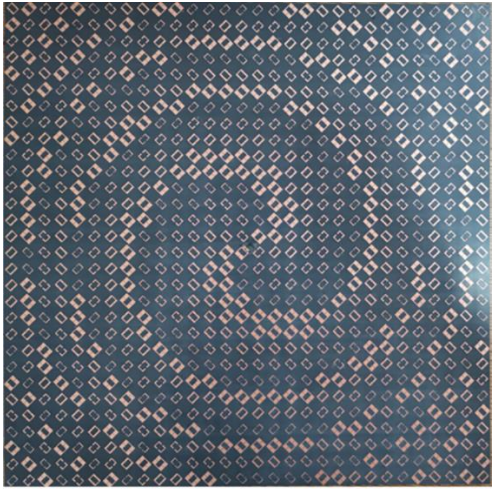

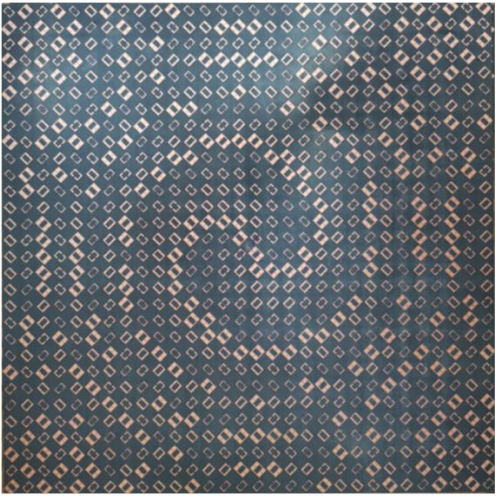

Fig. 8 (a) Experimental system configuration for the OAM vortex waves measurement in *x-y* plane. (b) Fabricated prototype of the OAM-generation reflectarray. (c) Fabricated prototype of the OAM-focusing reflectarray.

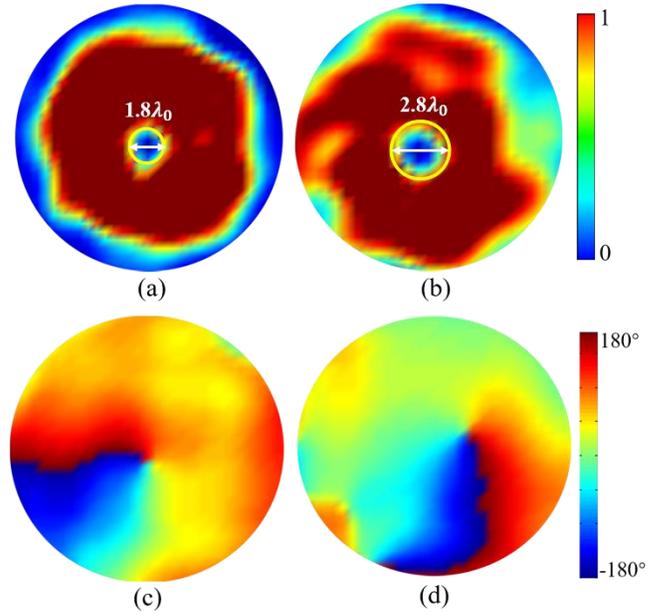

Fig. 9 Experimental results of OAM-generation vortex waves with topological charge *l* = 1 at 12 GHz in *x-y* plane. (a) Energy distribution of electrical field at cross sections *z* = 600 mm. (b) Energy distribution of electrical field at cross sections *z* = 800 mm. (c) Phase distribution of electrical field at cross sections *z* = 600 mm. (d) Phase distribution of electrical field at cross sections *z* = 800 mm.

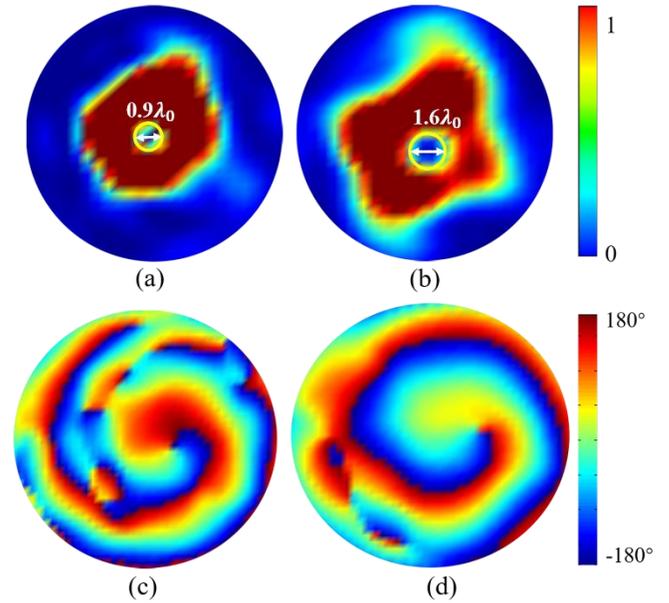

Fig. 10 Experimental results of the polarized reflectarray for OAM-focusing vortex beam with topological charge *l*=1 in *x-y* plane. (a) Energy distribution of electrical field at cross sections *z* = 600 mm. (b) Energy distribution of electrical field at cross sections *z* = 800 mm. (c) Phase distribution of electrical field at cross sections *z* = 600 mm. (d) Phase distribution of electrical field at cross sections *z* = 800 mm.

## V. Conclusion

In summary, we have proposed an approach to generate and focus OAM modes using polarized reflectarrays. These OAM-generation and focusing vortex waves with same topological charges have been theoretically and experimentally demonstrated in the microwave regime via phase analysis. Thus, based on the superposition of phase profile of spiral phase plate and focusing factor, the OAM-focusing beams can be achieved, their inner radius along the propagation direction being reduced as compared to that of OAM-generating reflectarray, the OAM-focusing reflectarray is obviously more constrictive, while the diameter of energy ring is about half of that OAM-generation reflectarray in the same position. The OAM-generating and focusing prototypes were fabricated and measured to verify the theoretical and numerical predictions. The agreement between the simulation and measurement results confirms the validity of our design procedure. By using the proposed configuration, it is much easier to produce the OAM-focusing vortex waves with different mode numbers, and it also opens a new way to generate and focus OAM vortex waves for microwave wireless communication applications. This proposed method would be particularly useful to design ultra-thin reflectarrays for focusing vortex beams that carry desirable OAM modes to potentially improve their communication efficiency.

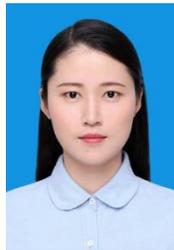

**Fengxia Li** received the B.S. degree in electronic information science and technology from Henan Polytechnic University, Jiaozuo, China, in 2015. She is currently pursuing the Ph.D. degree in microelectronics and solid states electronics at the University of Electronic Science and Technology of China, Chengdu, China.

Her recent research interests include the structural design of novel metamaterials and their applications in electromagnetic wave modulation, frequency selective surfaces, RCS reduction and wave absorbers.

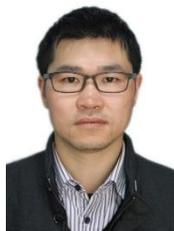

**Haiyan Chen** received the PH.D. degree in microelectronics and solid states electronics from the University of Electronic Science and Technology of China (UESTC), Chengdu, China, in 2011.

In 2015, he joined the Department of Engineering, University of Kentucky, Lexington, KY, USA, as a Visiting Scholar. Since 2012, he has been with the National Engineering Research Center of Electromagnetic Radiation Control Materials, UESTC, where he is currently a Professor. His current research interests include artificial electromagnetic (EM) material and EM radiation control materials, particularly study on EM discontinuous repair materials.

Dr. Chen was a recipient of the Chinese Scholarship Council Scholarship in 2015.

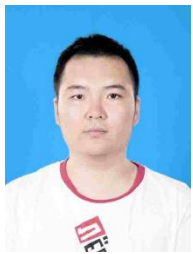

**Yang Zhou** was born in Liaoning, China, in 1989. He received the B.S. degree in electronic science and technology from the University of Electronic Science and Technology of China, Chengdu, China, in 2012, and the Ph.D. degree from the University of Electronic Science and Technology of China, in 2019.

His recently research activities have focused on the numerical modeling of novel metamaterials and their applications in wave modulation, antenna arrays, and wave absorber.

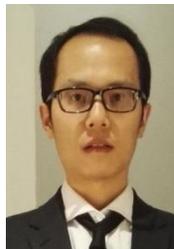

**Jian Wei You** received the B.Sc. degree in electrical engineering from Xidian University, Xi'an, China, in 2010, and the Ph.D. degree in electromagnetic and microwave technology from Southeast University, Nanjing, China, in 2016.

From July 2011 to December 2011, he was as a Research Visitor with the Department of Electrical and Computer Engineering, University of Houston, Houston, USA. In January 2016, he joined the Department of Electronic and Electrical Engineering, University College London, as a Research Associate.

His research interests include computational electromagnetics, nonlinear microwave and optics, quantum metamaterials, multiphysics, and computational plasma physics.

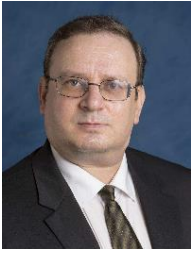

**Nicolae C. Panoiu** received the B.Sc. and M.S. degrees in physics from the University of Bucharest, Bucharest, Romania, in 1990 and 1992, respectively, and the Ph.D. degree from New York University (NYU), New York City, in 2001.

After graduating from NYU, he was a Postdoctoral Fellow with the Department of Applied Physics and Applied Mathematics, Columbia University, New York City. He is currently a Professor of Nanophotonics with the Department of Electronic and Electrical Engineering, University College London.

His research interests include silicon photonics, optical properties of photonic nanostructures and metamaterials, and computational modeling of electromagnetic structures. He is a member of OSA.

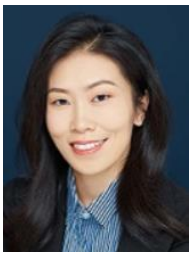

**Peiheng Zhou** received the Ph.D. degree in Material Physics and Chemistry from the University of Electronic Science and Technology of China (UESTC), Chengdu, China, in 2009.

She has been with the State Key Laboratory of Electronic Thin Films and Integrated Devices and the National Engineering Research Center of Electromagnetic Radiation Control Materials, UESTC, where she is currently a Full Professor. From November 2007 to November 2008, she was as a Research Visitor with the Massachusetts Institute of Technology, Cambridge, USA. From September 2018 to April 2020, she was as a Research Visitor with the Nanyang Technological University, Singapore.

Her current research interest includes various aspects of electromagnetic materials or structures, particularly the interplay between electromagnetic wave and materials.

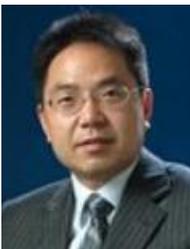

**Longjiang Deng** received the M.S. degree in electronic material and device from the University of Electronic Science and Technology of China (UESTC), Chengdu, China, in 1987.

Since then he has been working with UESTC, as a Lecturer, Associate Professor, and a Full Professor. He has authored/coauthored about 200 papers in refereed international journals and industry publications, and given many invited talks in international conferences. His research interests include electromagnetic wave absorbing material, infrared low emissivity and selective emissivity thin film, and microwave magnetic material.

Mr. Deng is a member of the Branch of Chinese Institute of Electronics on Microwave Magnetism, the Editor Committee of the Chinese Journal of Functional Materials, and the Vice Director of the Special Committee of the Chinese Institute of Electromagnetic Material.